\documentstyle{mn}
\newcommand{\reference}{\bibitem}
\input epsf

\def\lesssim{\mathrel{\hbox{\rlap{\hbox{\lower4pt\hbox{$\sim$}}}\hbox{$<$}}}}
\def\gtrsim{\mathrel{\hbox{\rlap{\hbox{\lower4pt\hbox{$\sim$}}}\hbox{$>$}}}}
\def\apj{ApJ}

\def\aj{AJ}
\def\aap{A\&\hskip-1pt A}

\def\mnras{MNRAS}

\def\nat{Nature}

\newcommand{\uvec}  {\mbox{\boldmath $u$}}
\newcommand{\xvec}  {\mbox{\boldmath $x$}}
\newcommand{\yvec}  {\mbox{\boldmath $y$}}
\newcommand{\deltavec}{\mbox{\boldmath $\delta$}}
\newcommand{\thetavec}  {\mbox{\boldmath $\theta$}}

\title[Astrometric Binary Source Effect]
      {The Effect of a Binary Source Companion
       on the Astrometric Microlensing Behavior}
\author[Han]
{
Cheongho Han\thanks{e-mails: cheongho@astroph.chungbuk.ac.kr}\\
Department of Physics, Chungbuk National University, Chongju 361-763, Korea
}

\begin{document}
\maketitle
\label{firstpage}
\vspace{-\abovedisplayskip} 
\begin{abstract}
If gravitational microlensing occurs in a binary-source system, both source 
components are magnified, and the resulting light curve deviates from the 
standard one of a single source event.  However, in most cases only one 
source component is highly magnified and the other component (the companion) 
can be treated as a simple blending source: blending approximation.  In 
this paper, we show that, unlike the light curves, the astrometric curves, 
representing the trajectories of the source image centroid, of an important 
fraction of binary-source events will not be sufficiently well modeled by 
the blending effect alone.  This is because the centroid shift induced by 
the source companion endures to considerable distances from the lens.  
Therefore, in determining the lens parameters from astrometric curves to 
be measured by future high-precision astrometric instruments, it will be 
important to take the full effect of the source companion into consideration.
\end{abstract}


\section{Introduction}
If microlensing occurs in a binary-source system, both source components 
are gravitationally magnified and the resulting light curve deviates from the 
standard one of a single source event (Griest \& Hu 1992; Han \& Gould 1997).  
However, the photometric binary-source effect becomes important only for 
the rare cases of the lens trajectory's passage close to both source stars 
and in most cases only one source component is significantly magnified.  
Then, the less magnified source component (the companion) can be treated 
simply as a blending source, and the resulting light curve may then be 
approximated by that of a blended single source event (Dominik 1998).

Although microlensing events have till now been observed only photometrically 
(Alcock et al.\ 1993; Aubourg et al.\ 1993, Udalski et al.\ 1993; Alard \& 
Guibert 1997; Abe et al.\ 1997), they could also be observed astrometrically 
by using high-precision instruments that will become available in the near 
future, e.g.\ the {\it Space Interferometry Mission} (SIM, Unwin et al.\ 
1998), and the interferometers to be mounted on the Keck (Colavita et al.\ 
1998) and VLT (Mariotti et al.\ 1998).  If an event is astrometrically 
observed, one can measure the displacements in the source star image centeroid 
with respect to its unlensed position.  Once the trajectory of the centroid 
shifts (the astrometric curve) is measured, the physical parameters (mass 
and location) of the lens can be better constrained (Miyamoto \& Yoshii 
1995; H\o\hskip-1pt g, Novikov \& Polnarev 1995; Walker 1995; Boden, Shao 
\& Van Buren 1998; Han \& Chang 1999).

The binary source companion also affects the astrometric lensing behavior.
However, whenever its astrometric effect was considered (Han \& Kim 1999; 
Dalal \& Griest 2001), it was always treated as a simple blending source 
due to the belief that the astrometric binary-source effect would be 
similar to the photometric one.  In this paper, we show that contrary to 
this belief the effect of the source companion on the astrometric lensing 
behaviors of a significant fraction of binary-source events cannot be 
sufficiently well described by the blending approximation, even when the 
corresponding light curves are well approximated by the blending effect 
alone.

The paper is organized as follows.  In \S\ 2, we describe the basics of 
binary source and blending effects on the light and astrometric curves of 
lensing events.  In \S\ 3, we the present astrometric curves and the 
corresponding light curves of example binary-source events to illustrate 
the relative difficulty in describing the astrometric lensing behaviors by 
the blending approximation alone.  In \S\ 4, we statistically estimate the 
range of the binary source separation and the companion light fraction 
where the blending approximation breaks down for the description of the 
photometric and astrometric lensing behaviors.  We conclude with \S\ 5.

\section{Binary source and Blending Effects}

\subsection{Binary Source Effect}

When a {\it single} source is lensed, its image is split into two.  The 
two images appear at the positions 
\begin{equation}
\thetavec_{{\rm I}\pm}={\theta_{\rm E}\over 2}
      \left( u\pm \sqrt{u^2+4} \right) \hat{\uvec}, 
\end{equation}
and have magnifications 
\begin{equation}
A_\pm = {1\over 2}\left( {u^2+2 \over u \sqrt{u^2+4}}\ \pm 1 \right), 
\end{equation}
where $\theta_{\rm E}$ is the angular Einstein ring radius, $\uvec$ is 
the lens-source separation vector normalized by $\theta_{\rm E}$, and 
$\hat{\uvec}$ represents its unit vector.  The angular Einstein ring 
radius is related to the physical parameters of the lens by
\begin{equation}
\theta_{\rm E} = \sqrt{4Gm\over c^2} \left( {1\over D_{\rm ol}}- 
{1\over D_{\rm os}}\right)^{1/2},
\end{equation}
where $m$ is the lens mass and $D_{\rm ol}$ and $D_{\rm os}$ represent 
the distances to the lens and source star, respectively.  The lens-source
separation vector is related to the lensing parameters by
\begin{equation}
\uvec = \left( {t-t_0\over t_{\rm E}}\right)\ \hat{\xvec}+\beta\ \hat{\yvec},
\end{equation}
where $t_{\rm E}$ represents the time required for the source to transit 
$\theta_{\rm E}$ (Einstein time scale), $\beta$ is the closest lens-source
separation in units of $\theta_{\rm E}$ (impact parameter), $t_0$ is the 
time at that moment, and the unit vectors $\hat{\xvec}$ and $\hat{\yvec}$
are parallel with  and normal to the direction of the lens-source motion.  
The two images formed by the lens cannot be resolved, but one can measure 
the total magnification and the shift of the light centroid of the
source star with respect to its unlensed position, which are represented 
respectively by  
\begin{equation}
A = A_+ + A_- = {u^2+2\over u\sqrt{u^2+4}},
\end{equation}
and
\begin{equation}
\deltavec={A_+\thetavec_{{\rm I}+} + A_-\thetavec_{{\rm I}-}\over A}-\uvec
         ={\uvec \over u^2+2}\theta_{\rm E}.
\end{equation}
The resulting light curve is symmetric with respect to $t_0$  and the 
trajectory of the centroid shift traces an ellipse with a semi-major axis 
$a=(\beta^2+2)^{-1/2}\theta_{\rm E}/2$ and an axis ratio $q=\beta
(\beta^2+2)^{-1/2}$ (Walker 1995; Jeong, Han \& Park 1999).

If lensing occurs in a binary-source system, on the other hand, both 
source components are gravitationally magnified.  For the binary-source 
event, the lensing behavior involved with each source can be treated as 
an independent single source event.  Then the observed light and 
astrometric curves are represented by 
\begin{equation}
F_{\rm BS} = A_1 F_1 + A_2 F_2, 
\end{equation}
and
\begin{equation}
\deltavec_{\rm BS} = 
{A_1 F_1(\uvec_1 + \deltavec_1) + 
                A_2 F_2(\uvec_2 + \deltavec_2) \over
	        A_1F_1 + A_2F_2} - 
               {F_1 \uvec_1 + F_2 \uvec_2 \over F_1+F_2},
\end{equation}
where $\uvec_i$ are the separation vectors between the lens and the 
individual source components, $A_i$ and $\deltavec_i$ are the magnifications 
and the centroid shifts of the individual single source events with baseline 
source fluxes $F_i$, and the subscripts $i=1$ and 2 are used to denote 
parameters related to the individual single source events ($i=1$ for the 
event with the higher magnification).  We note that the reference position 
of the centroid shift measurements for the binary-source event is not the 
location of the primary, i.e. $\uvec_1$, but the center of light between 
the unlensed source components, i.e. the second term in eq.\ (8).  In 
Figure 1, we present the lens system geometry of an example binary-source 
event.

\begin{figure}
\epsfysize=9cm
\centerline{\epsfbox{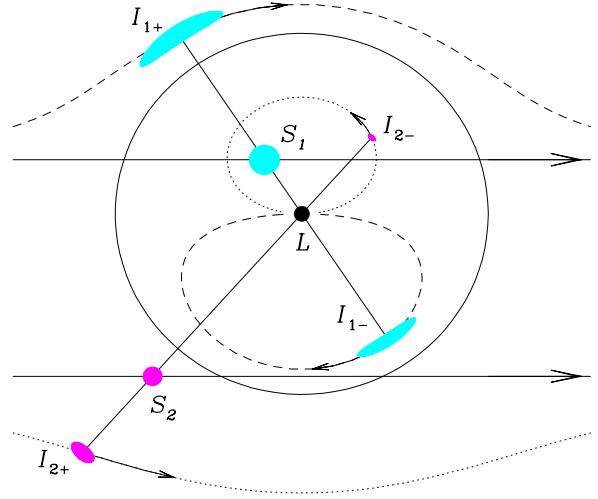}}
\vskip-1cm
\caption{
The lens system geometry of an example binary-source event.  The lens
(denoted by `{\it L}') is located at the center of the Einstein ring 
(large circle).  The two small filled circles (denoted by `$S_i$', where 
$i=1$ and 2) represent the locations of the two source stars at a specific 
moment.  The two distorted figures (denoted by `$I_{i+}$' and `$I_{i-}$') 
located along the line connecting each source and the lens are the images 
corresponding to the individual source stars.  The straight lines with 
arrows represent the source trajectories.  The curve passing through each 
image and the arrow on it represent the trajectory and direction of the 
image motion.
}
\end{figure}

\begin{figure*}
\epsfysize=15cm
\centerline{\epsfbox{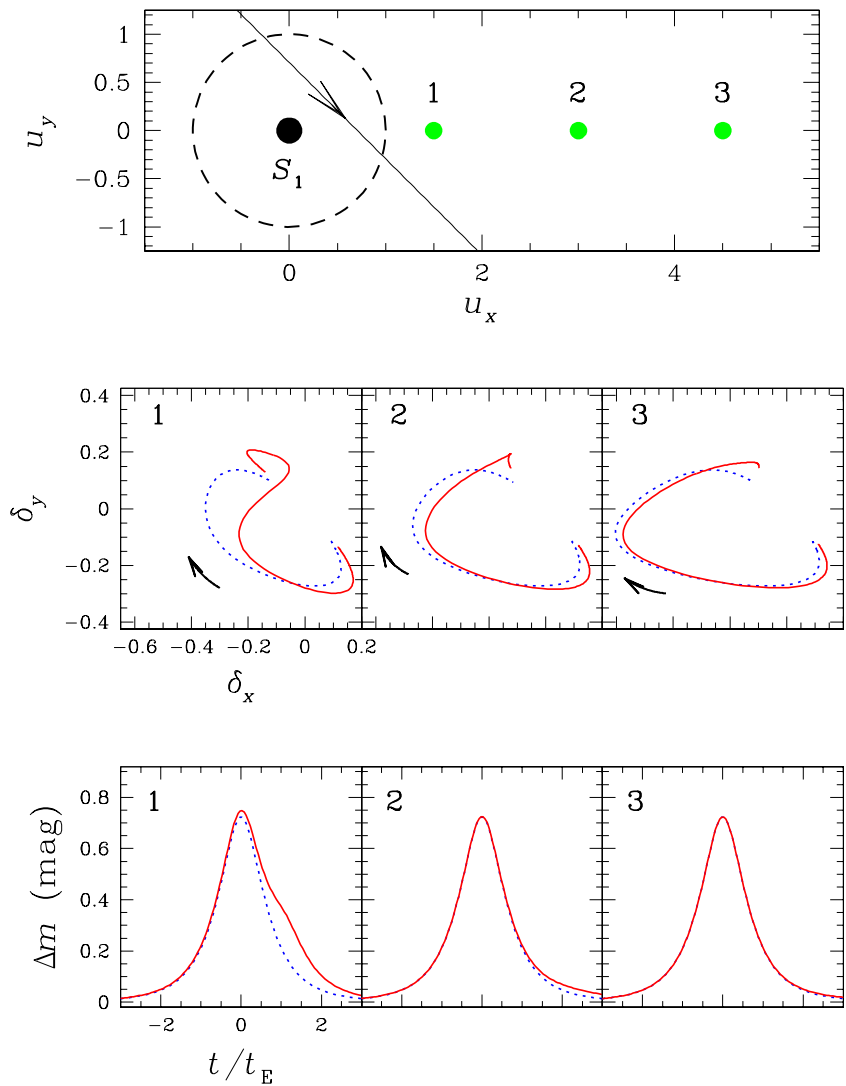}}
\caption{
Astrometric and light curves of binary-source events and in comparison with
those obtained by the blending approximation.  The upper panel shows the
geometry of the lens systems, where the primary (denoted by `$S_1$') is
located at the origin and the dots (designated by numbers) on the right 
side are the three different locations of the companion.  The straight line
with an arrow represents the lens trajectory.  The dashed circle represents
the lens locations where the magnification of the primary source is
$A_1=3/\sqrt{5}\sim 1.34$.  The middle and lower panels show how the  
astrometric and light curves change with the increasing source separation.
The number on each panel corresponds to that designating the companion 
location.  The solid curves are the exact astrometric and light curves of 
the binary-source events, and the dotted curves are those obtained by the 
blending approximation.  The companion has a light fraction of $f=0.2$.
}
\end{figure*}

\subsection{Blending Effect}

If a single source event is affected by the blended light from an unresolved 
nearby star, the observed light curve is represented by
\begin{equation}
F_{\rm blend} = A F_0 + F_{\rm b},
\end{equation}
where $F_0$ is the unblended baseline source flux and $F_{\rm b}$ is the 
amount of blended flux.  Since $F_{\rm b}$ is constant throughout the event, 
the observed flux is uniformly increased by $F_{\rm b}$ and the shape of the 
light curve is still symmetric.

The centroid shift of a blended event is represented by
\begin{equation}
\deltavec_{\rm blend} = {AF_0 (\uvec + \deltavec) + F_b \uvec_{\rm b} \over
	     AF_0 + F_{\rm b}} -
	    {F_0 \uvec + F_{\rm b} \uvec_{\rm b}\over
	     F_0 + F_{\rm b}},
\end{equation}
where $\uvec_{\rm b}$ is the position vector of the blending source with 
respect to the lens (Han \& Kim 1999; Dalal \& Griest 2001).  Blending 
has two effects on the observed astrometric curve.  Firstly, it causes the 
source star image centroid to be additionally shifted towards the blending 
source during the event.  Secondly, due to blending, the reference position 
of the centroid shift measurements is no longer the position of the lensed 
source but the center of light between the lensed star and the blending 
source.  As a result, depending on the position and the light fraction of 
the blending source, $f=F_{\rm b}/(F_0+F_{\rm b})$, the resulting astrometric 
curves can be seriously distorted from the elliptical one of the unblended 
event (Han \& Kim 1999; Dalal \& Griest 2001).

\begin{figure*}
\epsfysize=15cm
\centerline{\epsfbox{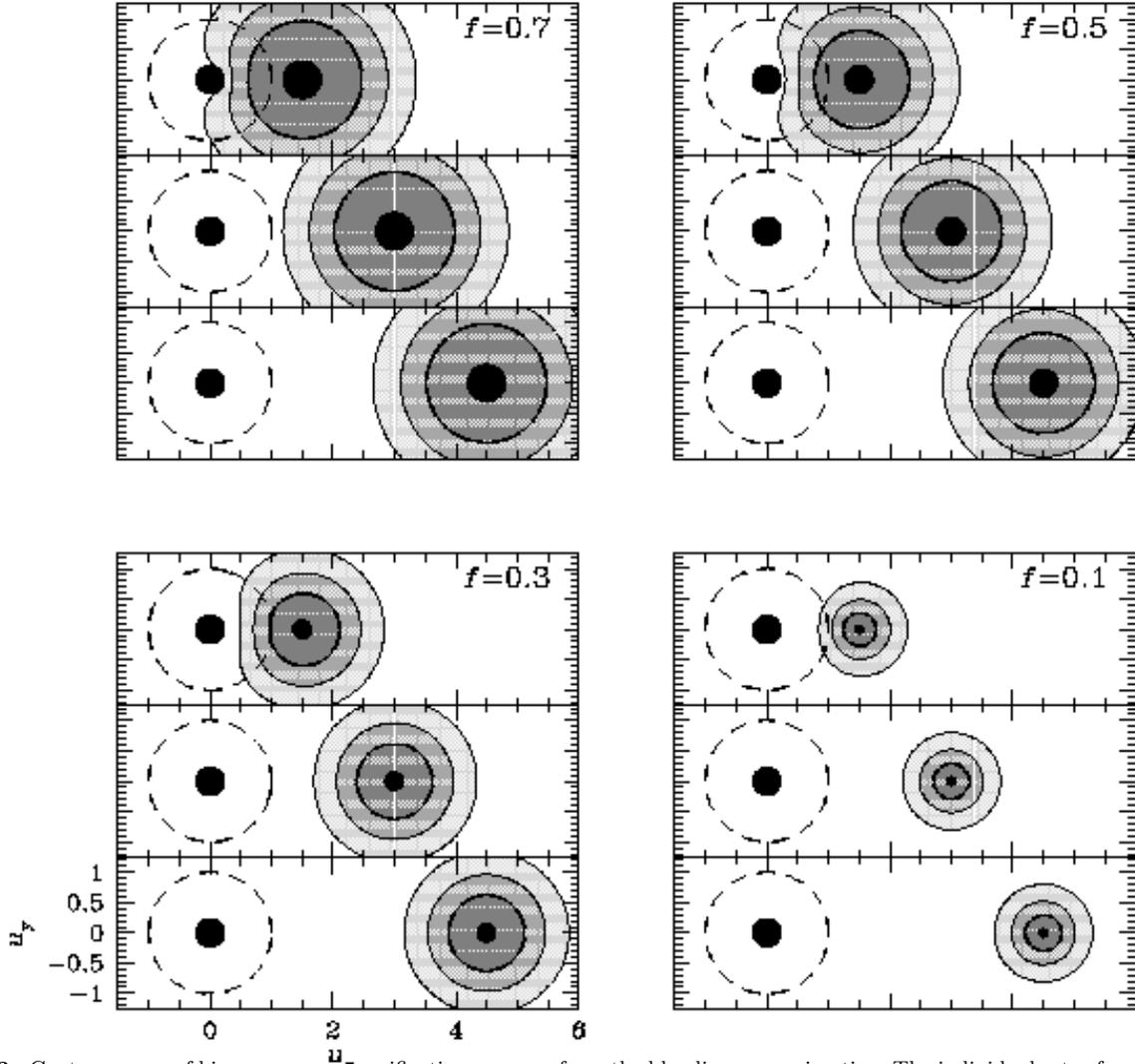}}
\vskip-0.8cm
\caption{
Contour maps of binary source magnification excesses from the blending 
approximation.  The individual sets of panels are the maps of binary source 
systems with different companion light fractions.  The panels in each set 
show the variation of the excess pattern with increasing source separation.
Contours are drawn at the levels of $\epsilon=5\%$, 10\%, and 20\%, 
respectively.
}
\end{figure*}

\begin{figure*}
\epsfysize=15cm
\centerline{\epsfbox{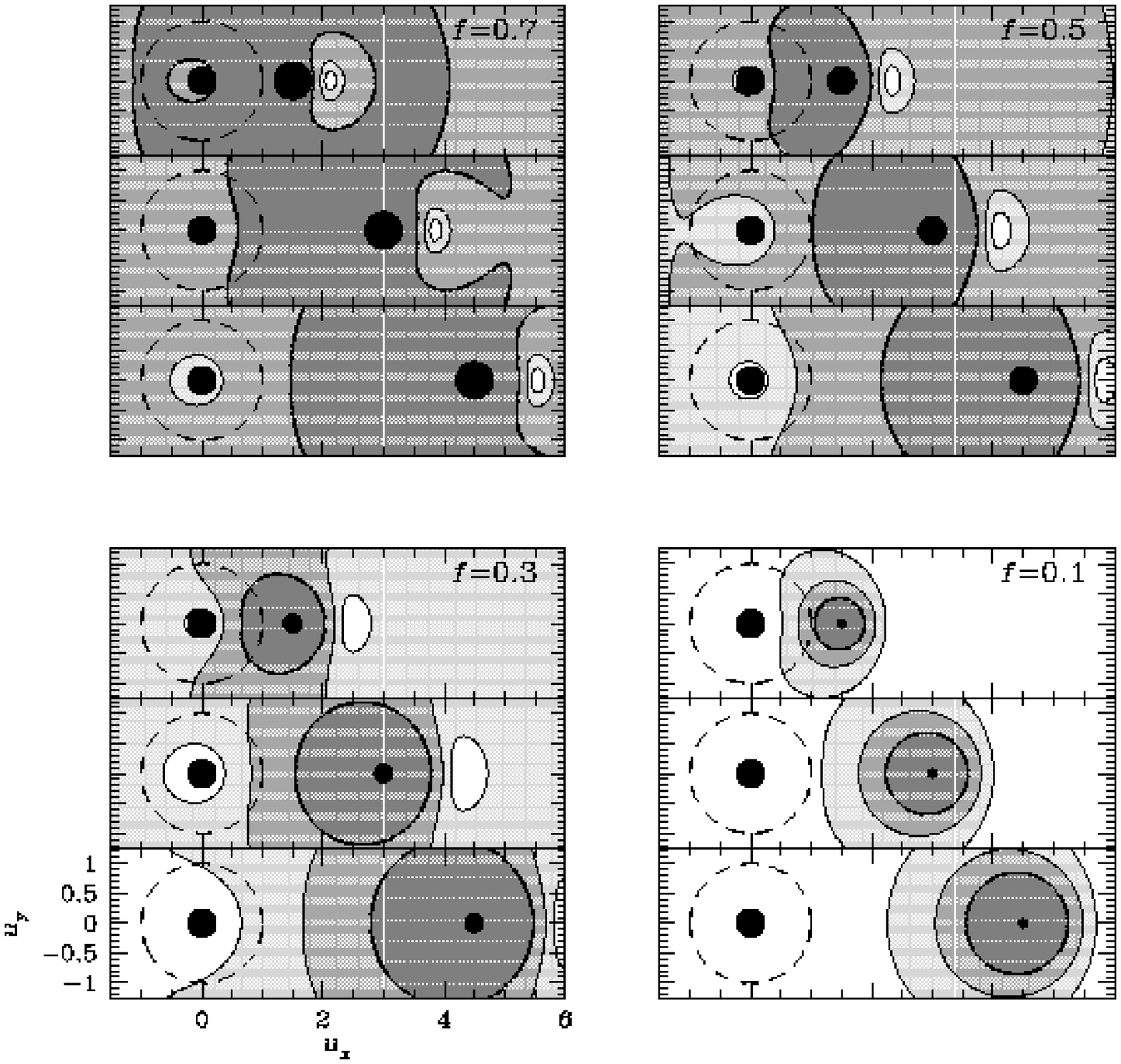}}
\vskip-0.8cm
\caption{
Similar maps as in Fig.\ 3, but for the astrometric centroid shift excesses
$\Delta\delta$.  Contours are drawn at the levels of $\Delta\delta=0.05
\theta_{\rm E}$, $0.1\theta_{\rm E}$, and $0.2\theta_{\rm E}$, respectively.
}
\end{figure*}

\section{Blending Approximation}

In many cases, the light curves of binary-source events resemble those of 
single source events.  This is because the lens, in general, approaches 
closely only one of the source components.  In this case, the magnification 
of the companion is small, i.e.\ $A_2\sim 1.0$, and thus the the resulting 
light curve can be approximated as 
\begin{equation}
A_{\rm BS} \sim A_1 F_1 + F_2.
\end{equation}
This implies that the observed light curve is well described by that of a 
single source event where the companion simply acts as a blending source.

Can the blending approximation be equally applied to astrometric curves 
of most binary-source events?  To this question, we find the answer is 
`no'.  We demonstrate this in Figure 2, where we present the astrometric 
(middle panels) and light curves (lower panels) of binary-source events 
with three different values of source separation, and we compare these 
curves to those obtained by the blending approximation.  One finds that 
even when the light curve is well approximated by the blending effect 
alone, the deviation of the astrometric curve from the approximation is 
considerable.

To see the patterns of both the photometric and astrometric deviations from 
the blending approximation for more general cases of binary-source events, 
we construct contour maps of magnification and centroid shift excesses for 
binary-source systems with various values of the binary separation, $\ell$, 
and the companion light fraction, $f$.  The magnification and centroid shift 
excesses are defined respectively by
\begin{equation}
\epsilon = {A_{\rm BS}-A_{\rm blend}\over A_{\rm blend}},
\end{equation}
and 
\begin{equation}
\Delta\deltavec = \deltavec_{\rm BS}-\deltavec_{\rm blend},
\end{equation}
where $A_{\rm BS}$ and $\deltavec_{\rm BS}$ are the exact magnification 
and centroid shift of the binary-source event, and $A_{\rm blend}$ and 
$\deltavec_{\rm blend}$ are those obtained by the blending approximation.
The constructed maps are presented in Figure 3 for the excess magnification,
and in Figure 4 for the excess centroid shift, respectively.  From a
comparison of the two maps, one finds that while significant photometric 
deviations occur only in a small region around the companion, comparable
astrometric deviations occur in a substantially larger area, even for a 
companion with a large separation and a small light fraction.

The relative difficulty in describing the astrometric curves of binary-source 
events by just the blending approximation can be understood in the following 
way.  If a source companion is located at a large distance from the lens 
(i.e.\ $u_{\rm 2}\gg\sqrt{2}$), its contribution to the magnification and 
the centroid shift are represented by
\begin{equation}
A_2 \sim 1 + {2\over u_2^4},
\end{equation}
and
\begin{equation}
\delta_2 \sim {\theta_{\rm E}\over u_2}.
\end{equation}
Then, as the separation becomes larger, the photometric contribution falls 
off rapidly ($\sim u_2^4$), while the astrometric contribution decays
much more slowly ($\sim u_2$) (Miralda-Escud\'e 1996; Paczy\'nski 1998).  
As a result, even at the location where the magnification of the companion 
is negligible (i.e.\ $A_2\sim 1$), the amount of the centroid shift can 
be considerable.  In this range of $u_{\rm 2}$, the centroid shift becomes
\begin{equation}
\deltavec_{\rm BS} \sim 
{A_1 F_1(\uvec_1 + \deltavec_1) + 
                F_2(\uvec_2 + \deltavec_2) \over
                A_1F_1 + F_2} - 
               {F_1 \uvec_1 + F_2 \uvec_2 \over F_1+F_2},
\end{equation}
which differs from $\deltavec_{\rm blend}$ in eq.\ (10).

\section{Validity of the Blending Approximation}

In the previous section, we illustrated the relative difficulties in 
describing the astrometric lensing behavior of binary-source events by 
the blending approximation, as compared to the photometric behavior.  
Then, a natural question to ask is: over what range of values of the source 
separation and the companion light fraction does the blending approximation 
break down for the description of the photometric and astrometric behaviors 
of binary-source events?  We answer this question as follows.

We proceed by statistically estimating the probabilities of binary-source 
events whose light and astrometric curves can be approximated by the 
blending effect as functions of $\ell$ and $f$; $P_{\rm ph} (\ell,f)$ for 
the photometric and $P_{\rm ast}(\ell,f)$ for the astrometric measurements.  
To determine these probabilities, we first simulate a large number of 
microlensing events occurred in binary-source systems, characterized 
by $\ell$ and $f$.  We then determine the probabilities by estimating 
the fraction of events whose light and astrometric curves have deviations 
(from those obtained by the blending approximation) less than preselected 
threshold values.  The lens trajectories of the tested events are selected 
so that they have random orientations with respect to the binary axis, and 
so that their impact parameters with respect to the primary source are 
uniformly distributed in the range of $0\leq \beta_1\leq 1$.  We assume that 
the blending approximation is valid if the light and astrometric curves, 
measured during the time period $-10 t_{\rm E}\leq t\leq 10 t_{\rm E}$, 
have deviations less than the threshold values of $\epsilon_{\rm th}=10\%$ 
and $\Delta\delta_{\rm th}=0.1\theta_{\rm E}$, respectively.

\begin{figure}
\epsfysize=11cm
\centerline{\epsfbox{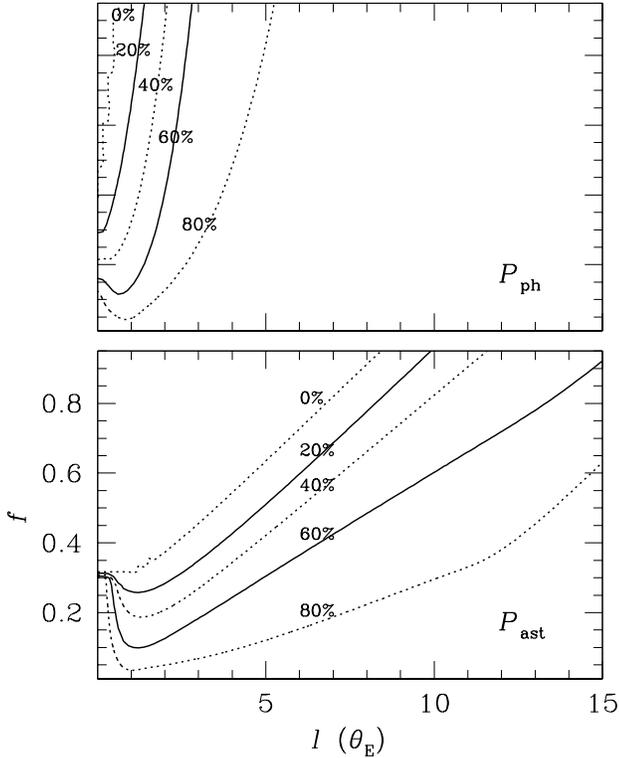}}
\caption{
The probabilities of binary-source events whose light and astrometric 
curves can be approximated by the blending well approximation, as 
functions of the binary separation, $\ell$, and the companion light 
fraction, $f$.  The upper and lower panels are for the photometric 
and astrometric probabilities, respectively.  
}
\end{figure}

In Figure 5, we present the resulting probabilities as contour maps in the 
parameter space of $\ell$ and $f$: upper panel for $P_{\rm ast}(\ell, f)$
and lower panel for $P_{\rm ast}(\ell, f)$.  From the figure, one finds 
that for a given $\ell$ and $f$, $P_{\rm ast}$ is significantly lower than 
the corresponding $P_{\rm ph}$.  For example, if microlensing occurs in a 
binary-source system with $\ell=6.0\theta_{\rm E}$, we estimate that the 
probabilities of events whose astrometric curves can be treated by the 
blending approximation will be only $P_{\rm ast}\sim 0\%$, 20\%, 50\%, and 
70\% for companions with light fractions of $f=0.8$, 0.6, 0.4, and 0.2, 
respectively; while the corresponding photometric probabilities will be 
$P_{\rm ph}\gtrsim 80\%$ for all four values of $f$.  One also finds 
that the region of low $P_{\rm ast}$ occupies a significantly larger area 
in the parameter space than the low $P_{\rm ph}$ region, implying that the 
astrometric blending approximation will not be valid for a significant 
fraction of binary-source events.

\section{Conclusion}

We have shown that although the blending approximation describes well
the photometric lensing behaviors of most binary-source events, the 
approximation will not be able to adequately describe the astrometric 
lensing behaviors for a significant fraction of binary-source events.  
This is because the astrometric effect of the companion endures to 
relatively large distances where the corresponding photometric effect 
is negligible.  Therefore, it will be important to take the full effect 
of the binary companion into consideration in determining the lens 
parameters from the observed source motion.

We thank Andr\'e Fletcher (Korea Astronomy Observatory) for his help with 
the preparation of the paper.  This work was supported by a grant 
(1999-2-1-13-001-5) from the Korea Science \& Engineering Foundation 
(KOSEF).

{}

\end{document}